\documentclass[journal=jpclcd,manuscript=article]{achemso}
\usepackage[version=3]{mhchem} % Formula subscripts using \ce{}
\usepackage{epstopdf}
\usepackage{graphicx}
\usepackage{color}

\author{Boyi Zhou}
\affiliation
{Department of Physics and Astronomy and Center for Simulational Physics, University of Georgia, Athens, Georgia 30602, United States}
\alsoaffiliation{Key Laboratory of Materials Modification by Laser, Electron, and lon Beams (Ministry of Education), School of Physics, Dalian University of Technology, Dalian 116024, P. R. China}
\author{Benhui Yang}
\affiliation
{Department of Physics and Astronomy and Center for Simulational Physics, University of Georgia, Athens, Georgia 30602, United States}
\author{N. Balakrishnan}
\affiliation{Department of Chemistry and Biochemistry, University of Nevada, Las Vegas, Nevada 89154, United States}
\author{B. K. Kendrick}
\affiliation{Theoretical Division (T-1, MS B221), Los Alamos National Laboratory, Los Alamos, New Mexico 87545, United States}
\author{P. C. Stancil}
\affiliation
{Department of Physics and Astronomy and Center for Simulational Physics, University of Georgia, Athens, Georgia 30602, United States}
\email{pstancil@uga.edu}
\title[An \textsf{achemso} demo]
  {Prediction of a Feshbach Resonance in the Below-the-Barrier Reactive Scattering of Vibrationally Excited HD with H}

\abbreviations{PES}

\begin{document}

%%%%%%%%%%%%%%%%%%%%%%%%%%%%%%%%%%%%%%%%%%%%%%%%%%%%%%%%%%%%%%%%%%%%%
%% The "tocentry" environment can be used to create an entry for the
%% graphical table of contents. It is given here as some journals
%% require that it is printed as part of the abstract page. It will
%% be automatically moved as appropriate.
%%%%%%%%%%%%%%%%%%%%%%%%%%%%%%%%%%%%%%%%%%%%%%%%%%%%%%%%%%%%%%%%%%%%%
\begin{tocentry}

\includegraphics[width=5cm, height=5cm]{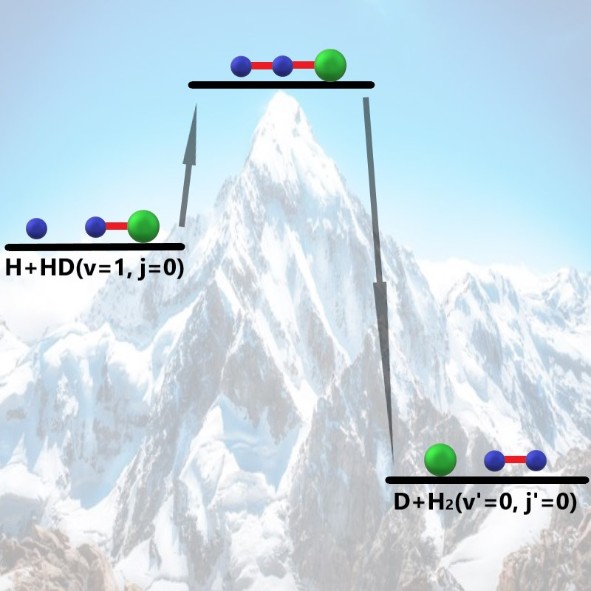}

\end{tocentry}

%%%%%%%%%%%%%%%%%%%%%%%%%%%%%%%%%%%%%%%%%%%%%%%%%%%%%%%%%%%%%%%%%%%%%
%% The abstract environment will automatically gobble the contents
%% if an abstract is not used by the target journal.
%%%%%%%%%%%%%%%%%%%%%%%%%%%%%%%%%%%%%%%%%%%%%%%%%%%%%%%%%%%%%%%%%%%%%
\begin{abstract}
Quantum reactive scattering calculations on the vibrational quenching of HD due to collisions with H were carried out employing an accurate potential energy surface. The state-to-state cross sections for the chemical reaction HD ($v=1, \ j=0$) + H $\rightarrow$ D + \ce{H2} ($v'=0, \ j'$) at collision energies between 1 and 10,000 cm$^{-1}$ are presented, and a Feshbach resonance in the low-energy regime, below the reaction barrier, is observed for the first time. The resonance is attributed to coupling with the vibrationally adiabatic potential correlating to the $v=1, \ j=1$ level of the HD molecule, and it is dominated by the contribution from a single partial wave. The properties of the resonance, such as its dynamic behavior, phase behavior, and lifetime, are discussed.
\end{abstract}

%%%%%%%%%%%%%%%%%%%%%%%%%%%%%%%%%%%%%%%%%%%%%%%%%%%%%%%%%%%%%%%%%%%%%
%% Start the main part of the manuscript here.
%%%%%%%%%%%%%%%%%%%%%%%%%%%%%%%%%%%%%%%%%%%%%%%%%%%%%%%%%%%%%%%%%%%%%
The study of reactive collisions in atom-diatom systems is a very effective way to probe the interactions and reaction dynamics between atoms and molecules. Therefore, triatomic collision processes have played an important role in quantum dynamics, as they can give valuable insight into the chemical reaction mechanism.\cite{yang2019enhanced,yang2020}
Because of their light mass and the availability of high-accuracy molecular potentials, \ce{H2}, HD, and \ce{D2} are the most commonly studied molecular species and they serve as benchmark systems for state-to-state chemistry. Because of their high abundance in the interstellar medium, collisions of \ce{H2} and HD with  atomic hydrogen have attracted significant interest. For practical reasons, D + \ce{H2} 
\cite{zhang1989quantum,takada1992reaction,aoiz1996reaction,banares1997quantum}
and H + \ce{D2} 
\cite{yuan2020imaging,marinero1984h+,rinnen1988h+,d1991quantum,gao2015differential}
reactions have been studied rather extensively. HD, as the singly deuterated form of molecular hydrogen, is one of the products in these reactions. Because of the relatively large abundance of HD in space and its finite dipole moment, the observable properties of the collisional reaction of HD with H, such as cross sections and rate coefficients, are of particular interest. 

Much attention has been devoted to the study of the H + HD reaction recently, stimulated by geometric phase effects in this system and the crucial role of HD in astrophysics.\cite{yuan2018direct,yuan2018observation,desrousseaux2018rotational,kendrick2016geometric1,kendrick2016geometric2,hazra2016geometric,kendrick2015geometric}
However, most of these studies, in particular those relevant to astrophysics, have investigated only rotational excitation of HD in its ground vibrational state. Vibrationally excited HD can be produced by a variety of processes in space, and thus, it is of interest to study its interaction with atomic hydrogen.\cite{kendrick2019nonadiabatic} Inspired by this point, we focus on the study of the reaction between vibrationally excited HD molecules and atomic hydrogen. The stereodynamics of rotational quenching of state-prepared HD in the $v=1$ vibrational level in collisions with \ce{H2}, \ce{D2}, and He have also been reported recently
\cite{perreault2017quantum,perreault2018cold,croft2018unraveling,croft2019controlling,perreault2019hd}. In this paper, we present results for the cross sections of the reactive processes for the H + HD collisional system when HD is in its first excited vibrational state ($v=1$) and discuss the behavior of vibrational quenching of HD molecules due to collisions with H atoms. In particular, we report the presence of a narrow Feshbach resonance in the HD + H $\rightarrow$ D + \ce{H2} reaction at energies below the reaction barrier that originates from coupling with the adiabatic potential correlating with the $v=1,j=1$ level of the HD molecule.

A reliable potential energy surface (PES) for the H + HD system ensures the accuracy of the quantum dynamics calculations. Thus, an accurate description of the PES is required in order to obtain the cross sections for the reaction of H with HD. There are several PESs for the \ce{H3} system that have been widely used in quantum-mechanical calculations. The best known are the LSTH,
\cite{liu1973ab,siegbahn1978accurate,truhlar1978functional}
BKMP2,
\cite{boothroyd1996refined}
and CCI
\cite{mielke2002hierarchical}
potentials. Among them, the BKMP2 and CCI potentials are considered the most accurate, and they reproduce most of the available experimental data. In our scattering calculations, we employed the CCI PES, but some comparisons are also provided on the BKMP2 PES. The CCI PES was developed with \textit{ab initio} calculations of nearly full configuration interaction quality using a highly accurate many-body basis set extrapolation. There is a large barrier of about 3500 cm$^{-1}$ on this PES for the hydrogen exchange reaction with ground-state reactants, and the well depth of the global minimum is about 20 cm$^{-1}$.

In order to describe the reactive scattering of HD with H, the quantum-mechanical reactive scattering program ABC
\cite{skouteris2000abc}
was employed. To solve the Schr\"{o}dinger equation, ABC uses a coupled-channel hyperspherical coordinate method for the motion of the three nuclei on a Born-Oppenheimer PES. The hyper-radius is divided into a large number of sectors, and in each sector the wave function is expanded in terms of ro-vibrational eigenfunctions of the diatomic fragments in each atom-diatom arrangement channel. For low collision energies, we used a large maximum hyper-radius and a large number of log derivative propagation sectors to ensure that the integration step size $\Delta \rho$ would be small enough to obtain converged results. With the convergence parameters used in ABC (see Table S1), the calculations were carried out for total angular momentum quantum numbers $J$ from 0 to 120 in the range of collision energies from 1 to 10000 cm$^{-1}$. The results of the ABC program are the elements of the S-matrix in parity-adapted form. To obtain the collision energy variation of the cross sections for reactive processes, the elements of the S-matrix need to be converted from the parity-adapted form \textit{S$_{n'k',nk}^{J,P}$(E)} into standard helicity-representation form \textit{S$_{n'k',nk}^{J}$(E)}.
\cite{skouteris2000abc}
After the conversion, the reactive scattering integral cross sections (ICSs) can be calculated as
\begin{equation}
   \sigma_{{n'k'}\leftarrow nk}(E)=\frac{\pi}{k_n^2}\sum_J(2J+1)|{S_{n'k',nk}^{J}(E)}|^2,
\end{equation}
where $J$ is the total angular momentum quantum number, $E$ is the collision energy, $n$ and $n'$ are composite indices for ${\alpha}vj$ and ${\alpha}'v'j'$ (in which ${\alpha}$ and ${\alpha}'$ are arrangement labels), $k$ and $k'$ are helicity quantum numbers, ${k_n}$ is the wave vector (with 
\begin{math}
{k_n^2}=\frac{2\mu E}{\hbar^2},
\end{math}
), and $\mu$ is the reduced mass of the system.

Figure 1 plots the state-to-state cross section as a function of collision energy for the reaction HD ($v=1, \ j=0$) + H $\rightarrow$ D + \ce{H2} ($v'=0, \ j'=0$). As can be seen, the cross section at low energies is small and decreases with increasing collision energy, reaching a global minimum  at 84.69 cm$^{-1}$. The broader hump appearing at around 750 cm$^{-1}$ can be interpreted as metastable states of the H + HD reaction.
\cite{der2002signatures} The system passes through a transition state region along the reaction coordinate, and the maximum energy along the minimum-energy reaction pathway for the H + HD reaction is about 1036 cm$^{-1}$. Therefore, the cross section increases rapidly above a collision energy of 1000 cm$^{-1}$. Below this barrier, the reactivity can be attributed to quantum tunneling.

\begin{figure}[htbp]
  \centering
  \includegraphics[width=0.7\textwidth]{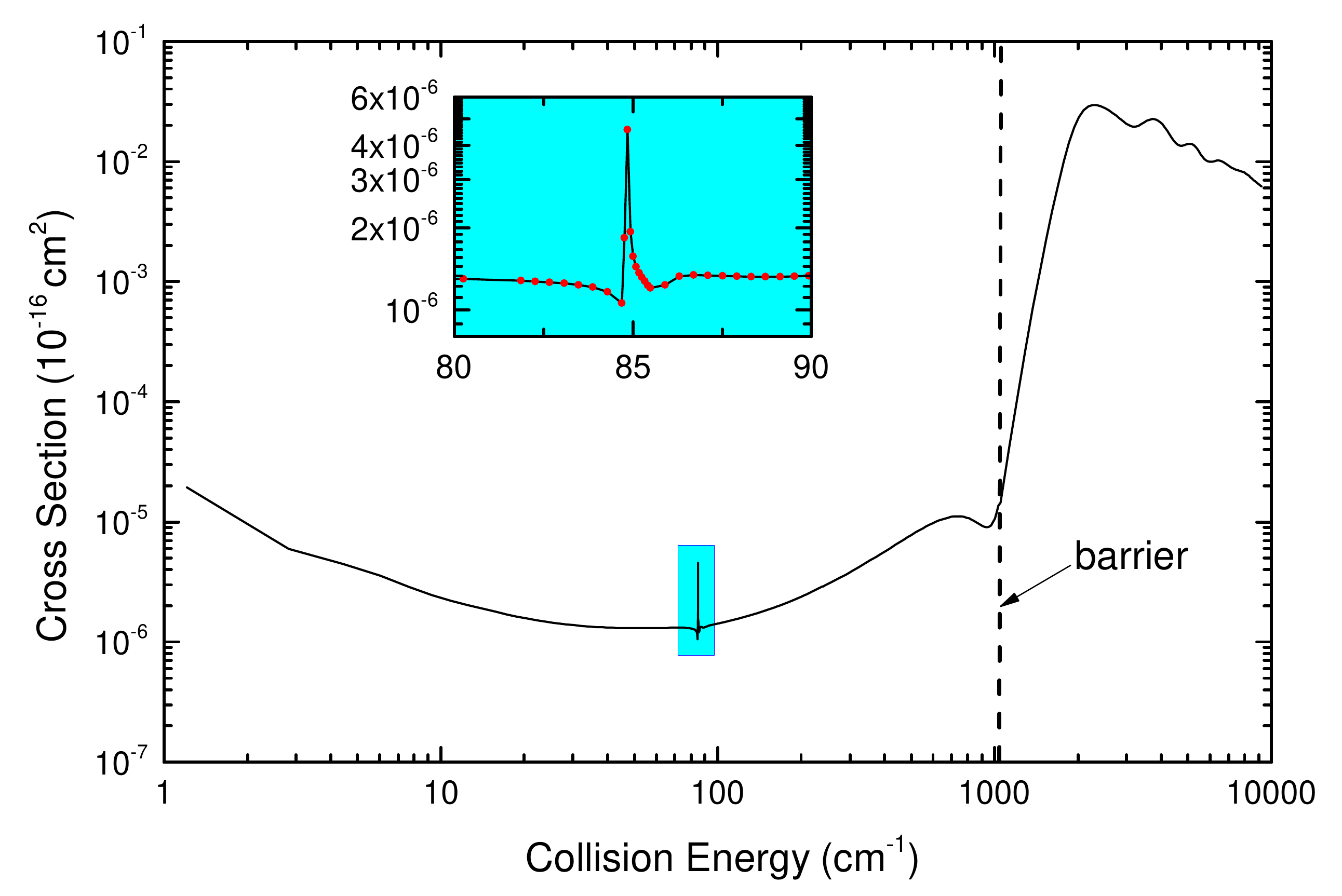}
  \caption{Collision energy variation of the cross section for the HD ($v=1, \ j=0$) + H $\rightarrow$ D + \ce{H2} ($v'=0, \ j'=0$) reaction. Inset: Feshbach resonance region.}
  \label{Cross_100_000}
\end{figure}

A striking feature of the results presented in Figure 1 is a sharp resonance peak that appears at 84.85 cm$^{-1}$, as shown in the inset. To our knowledge, this is the first time a resonance for the \ce{H3} system has been predicted in this energy range and below the reaction barrier. In order to confirm this observation, we performed the calculation of state-to-state cross sections with different final states. Figure 2 presents the collision energy variation of cross sections for the HD ($v=1, \ j=0$) + H $\rightarrow$ D + \ce{H2} ($v'=0, \ j'$) reaction with $j'$ from 1 to 4. All of the resonance peaks are located at 84.85 cm$^{-1}$, but they vary in magnitude. The resonance peaks decrease in going from $j'=$1 to 4, but it is worth noting that the peaks for $j'=1$ and $j'=2$ are larger than that for $j'=0$. We also computed the total integral cross sections summed over all final states. As can be seen in Figure S1, the resonance survives in the total vibrationally resolved cross sections.

\begin{figure}[htbp]
  \centering
  \includegraphics[width=0.7\textwidth]{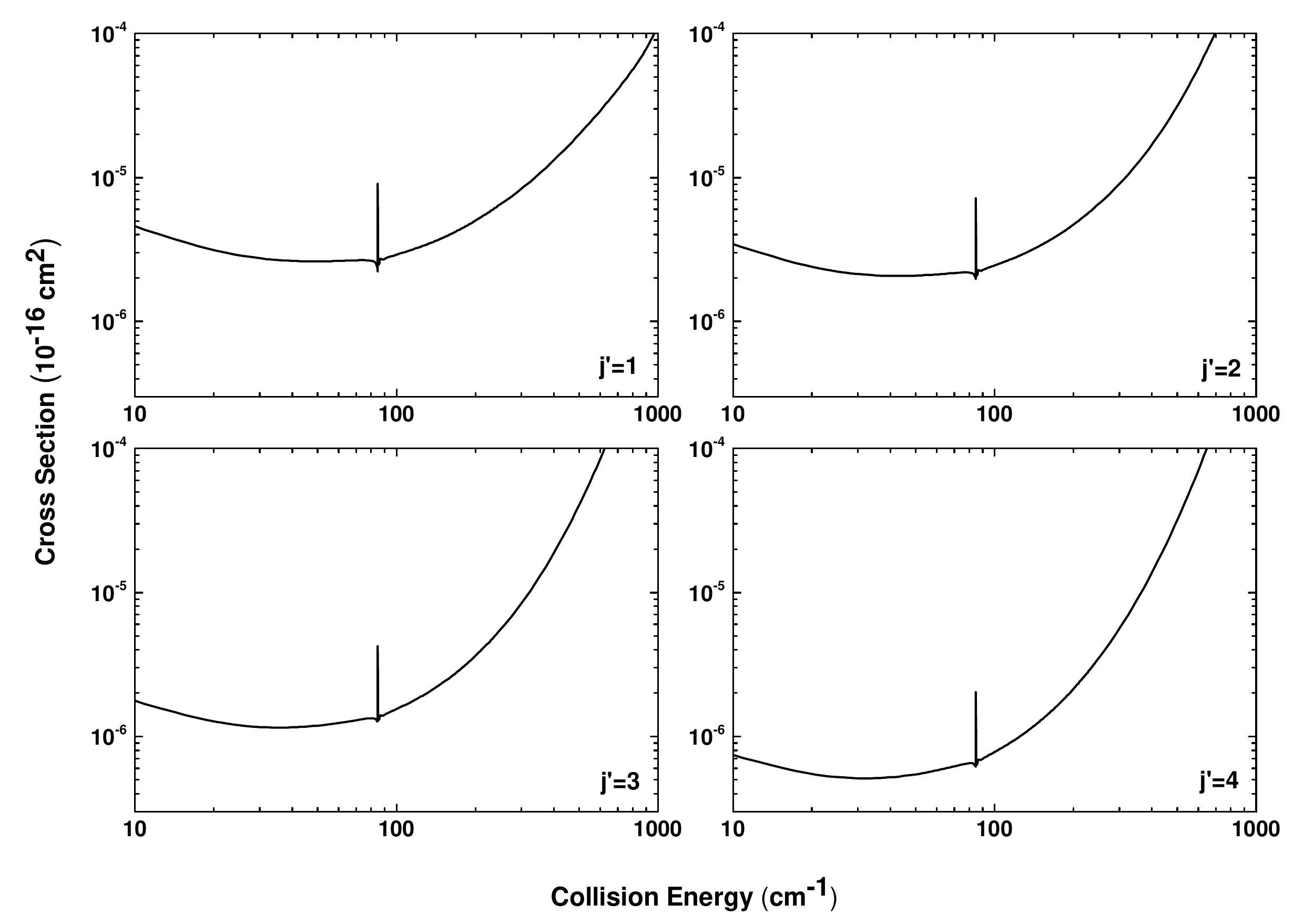}
  \caption{Collision energy variation of the cross sections for the HD ($v=1, \ j=0$) + H $\rightarrow$ D + \ce{H2} ($v'=0, \ j'$) reaction with $j'$ = 1, 2, 3, and 4.}
  \label{Different final states}
\end{figure}

Because resonance peaks are usually dominated by the contribution from a single partial waves or a narrow range of partial waves, the squares of the $S$-matrix elements \textit{S$_{n'k',nk}^{J}$(E)} for $J=0-6$ are plotted in Figure 3 (we note that here the partial wave or orbital angular momentum $L$ of H about HD is equal to $J$ since $j=0$). It is clear that $J=1$ dominates the peak at 84.85 cm$^{-1}$. The behavior that $J=1$ dominates the resonance can also be seen in the cross sections for $j'$ from 1 to 3 (see Figure S2). As shown in Figure 3, the broader hump in the region of at about 750 cm$^{-1}$ can be attributed to a nearly in-phase sum of $J$-resolved cross sections. To confirm that this resonance feature is not an artifact of the CCI PES or the ABC code used for the scattering  calculations, separate quantum mechanical reactive scattering calculations were also carried out on the BKMP2 potential using the APH3D quantum mechanical reactive scattering program.
\cite{kendrick1995recombination, kendrick2003quantum}
Indeed, the BKMP2 PES also captures this resonance at exactly the same energy, though the magnitudes of the background cross sections from the two calculations differ at energies below the barrier where tunneling contributions become very sensitive to the details of the PESs and the basis sets employed in the ABC and APH codes (see Figure S3). 

\begin{figure}[htbp]
  \centering
  \includegraphics[width=0.7\textwidth]{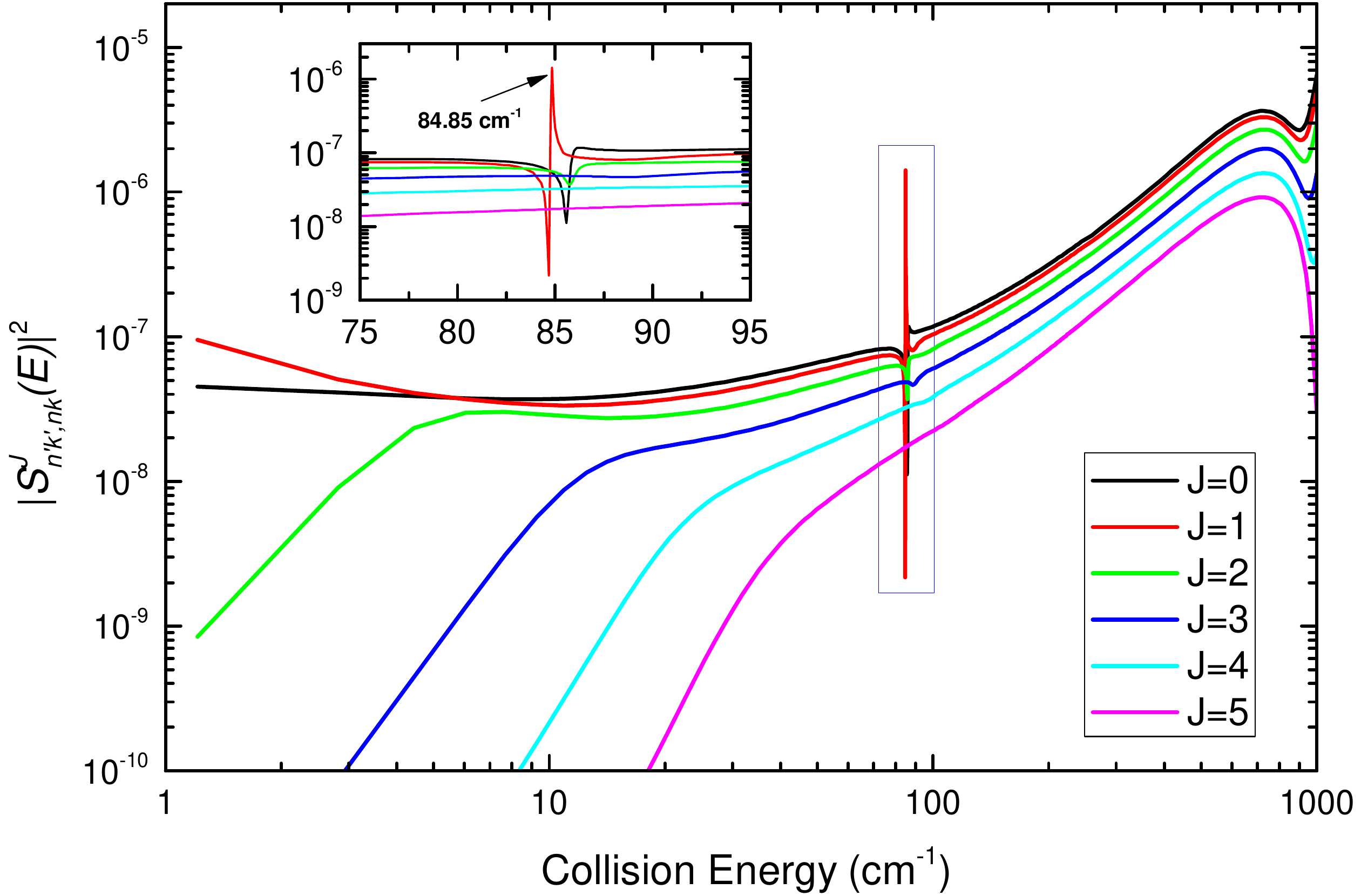}
  \caption{Collision energy variation of $|{S_{n'k',nk}^{J}(E)}|^2$ for the HD ($v=1, \ j=0$) + H $\rightarrow$ D + \ce{H2} ($v'=0, \ j'=0$) reaction for different values of the total angular momentum quantum number $J$.}
  \label{Partial Waves}
\end{figure}

\begin{figure}[htbp]
  \centering
  \includegraphics[width=0.5\textwidth]{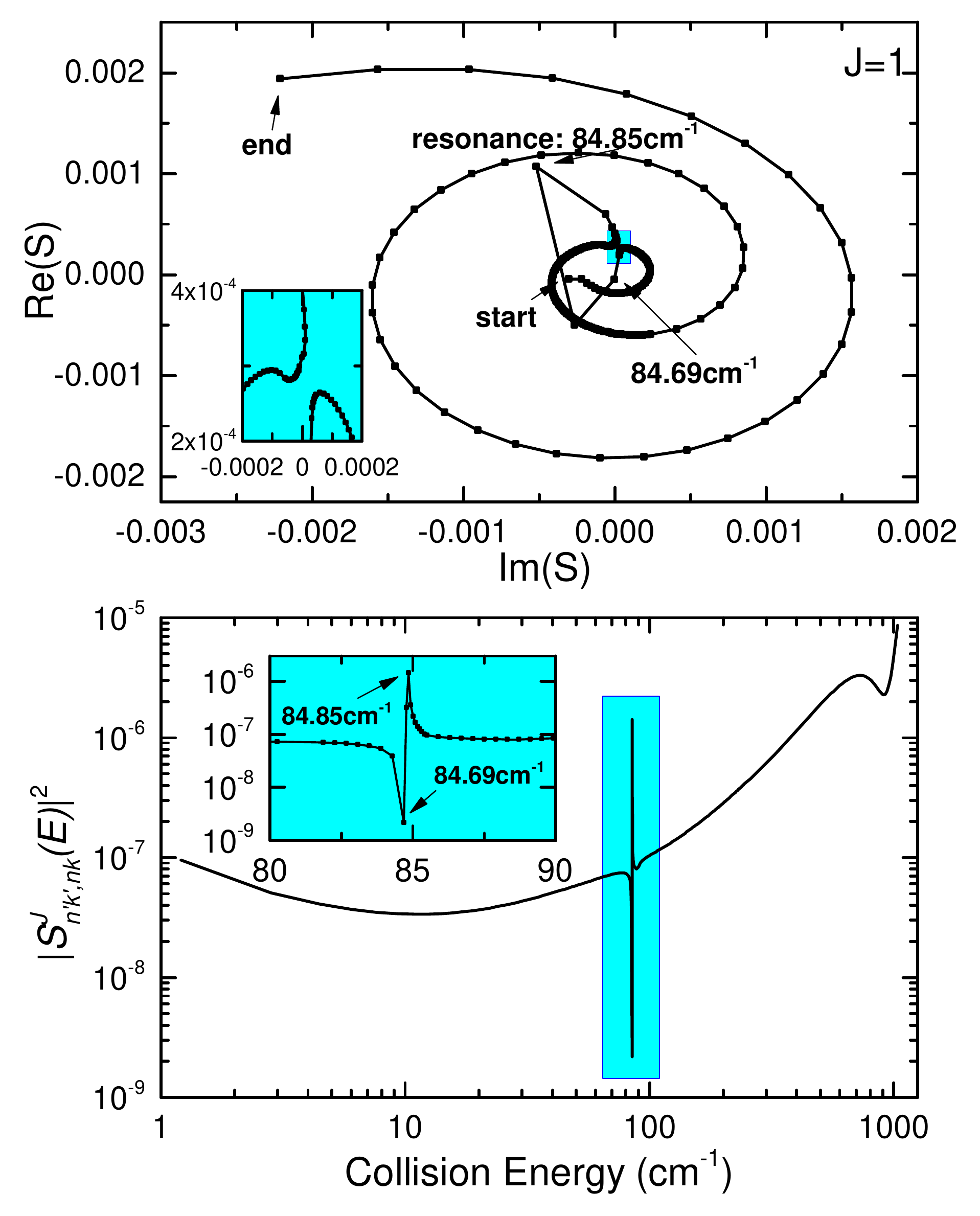}
  \caption{(top) Argand diagram for the HD ($v=1, \ j=0$) + H $\rightarrow$ D + \ce{H2} ($v'=0, \ j'=0$) reaction for $J$ = 1. (bottom) Corresponding values of $|{S_{n'k',nk}^{J}(E)}|^2$.}
  \label{Argand Plot}
\end{figure}

To confirm that the resonance exhibits the proper phase behavior,  an Argand diagram for the HD ($v=1, \ j=0$) + H $\rightarrow$ D + \ce{H2} ($v'=0, \ j'=0$) reaction for $J=1$ is presented in Figure 4. The collision energy increases from 1.2 to 1033.6 cm$^{-1}$. The points were computed on a fine energy grid near the resonance to display its phase behavior. The background phase behavior corresponds to the smooth counterclockwise motion of the energy trajectory in the Argand plot. Near the resonance peak (blue inset), a kink in the Argand plot is observed where the resonance phase shift dominates and the trajectory temporarily reverses direction and exhibits a clockwise motion. The kink and associated clockwise loop in the Argand plot are a signature of a quantum resonance. Thus, the observation of the peak in the cross section is reconfirmed from the phase behavior. Specifically, the resonance point that we note in the Argand diagram corresponds to the sharp peak located at 84.85 cm$^{-1}$, and the contribution at 84.69 cm$^{-1}$ causes the global minimum of the cross section shown in Figure 1 at the same collision energy.

To gain more insight into the origin of the resonance, we examined the adiabatic potentials for the H + HD interaction for $J=1$. Because the resonance occurs at about 85 cm$^{-1}$ relative to the $v=1, \ j=0$ level, it is most likely due to coupling with $v=1, \ j=1$ level of HD, which becomes energetically accessible at a collision energy of 85.37 cm$^{-1}$. Thus, it is very likely that this is a Feshbach resonance due to coupling with a quasi-bound state of the adiabatic potential correlating with the  $v=1, \ j=1$ level of HD. For the initial $v=1, \ j=0$ state of HD only $L=1$ contributes for $J=1$. Thus, only odd inversion parity  ($P=(-1)^{j + L} = -1$) contributes. For the $v=1, \ j=1$ level, odd-inversion-parity states for $J=1$ that couple to the $v=1, j=0, L=1$ channel are limited to $L=0$ and 2. The corresponding adiabatic potentials are shown in Figure 5 as functions of the center-of-mass separation $R$ between H and HD. The adiabatic potentials were computed by diagonalizing the interaction potential matrix  at each value of $R$. Since the resonance occurs at the same energy for different $j'$ levels of the product H$_2$, this is an entrance channel feature and we used reactant Jacobi coordinates and included only ro-vibrational levels of HD in constructing the matrix elements of the interaction potential. It can be seen that the adiabatic potential corresponding to $v=1, \ j=1, \ L=0$ supports a quasi-bound state at an energy of 84.82 cm$^{-1}$, which is very close to the energy of the resonance. The $v=1, \ j=1, \ L=2$ potential does not support a bound state. We carried out similar analysis with the BMKP2 potential, which also supports a bound state for the $v=1, \ j=1, \ L=0$ potential at 84.98 cm$^{-1}$ (see Figure S4), consistent with a similar resonance feature at 85.33 cm$^{-1}$ shown in Figure S3.

\begin{figure}[htbp]
  \centering 
  \includegraphics[width=0.7\textwidth]{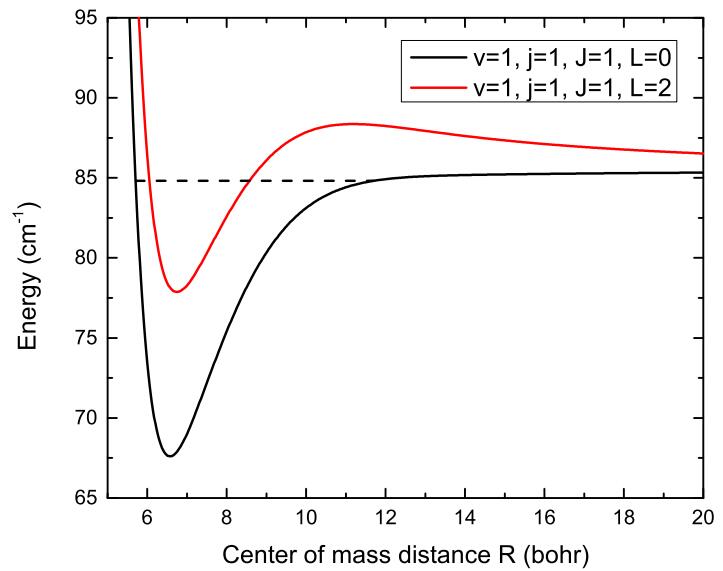}
  \caption{Adiabatic potentials correlating with different ($v, \ j$) states of HD in H + HD ($v=1, \ j=0$) collisions for $J=1$ on the CCI PES.}
  \label{Bound States}
\end{figure}

Resonances are often associated with metastable states of lifetime $\tau \simeq \hbar/\Gamma$, where $\Gamma$ is the width of the resonance. To guide future experiments, we carried out a Lorentzian fit and extracted the lifetime. A very high resolution energy scan in the resonance region using Smith's collision lifetime matrix formulation 
\cite{smith1960lifetime} 
was performed using the APH3D program on the CCI (BKMP2) PES. The results show that the lifetime is about 0.6 s (0.2 s) for this resonance state (see Figures S5 and S6). As can be seen, the lifetimes on the two PESs are very similar, and there is a slight shift in resonance energy: 85.37 cm$^{-1}$ for CCI versus 85.33 cm$^{-1}$ for BKMP2.

From an experimental point of view, a crossed-molecular-beam approach may be a good method to verify our theoretical prediction. A variety of experimental techniques have been developed over the past decade including, for example, Stark-induced adiabatic Raman passage (SARP),
\cite{mukherjee2018stark}
 the H-atom Rydberg tagging technique (HRTOF),
\cite{harich2002forward,dai2003interference}
 the photoinitiated reaction method (PHOTOLOC),
\cite{gao2015differential,jankunas2012seemingly,jankunas2014simplest}
 the velocity map ion imaging (VMI) technique,
\cite{eppink1997velocity,lin2003application}
resonance-enhanced multiphoton ionization (REMPI) with UV pulses,
\cite{zimmermann2008ultracompact} and the quantum-state-specific backward scattering spectroscopy (QSSBSS) method. \cite{yang2019enhanced} SARP can be used to prepare HD molecules in a specific internal state. The HRTOF, PHOTOLOG, REMPI with UV pulses, and QSSBSS approaches can be used for detection. The Feshbach resonance is predicted to appear in a low-collision-energy region, be very narrow, and have a small magnitude. Therefore, an experimental approach will require very fine kinetic energy resolution and sensitivity.

In conclusion, quantum reactive scattering calculations for the H + HD ($v=1, \ j=0$) reaction were performed to obtain the collision energy variation of the cross section. A Feshbach resonance due to coupling with the $v=1, \ j=1, \ L=0$ potential was found at about 85 cm$^{-1}$, which is significantly below the $\sim$1000 cm$^{-1}$ reaction barrier. In this reaction, the peak is dominated by the partial wave $L=1$. The resonance is very longlived, with a lifetime close to a tenth of a second. We hope that these results stimulate experimental exploration of this resonance and further in-depth theoretical studies of this reaction.

%%%%%%%%%%%%%%%%%%%%%%%%%%%%%%%%%%%%%%%%%%%%%%%%%%%%%%%%%%%%%%%%%%%%%
%% The "Acknowledgement" section can be given in all manuscript
%% classes.  This should be given within the "acknowledgement"
%% environment, which will make the correct section or running title.
%%%%%%%%%%%%%%%%%%%%%%%%%%%%%%%%%%%%%%%%%%%%%%%%%%%%%%%%%%%%%%%%%%%%%
\begin{acknowledgement}
Work at UGA was supported by NASA grant NNX16AF09G and at UNLV by NSF Grant No. PHY-1806334. The
China Scholarship Council is acknowledged for supporting Boyi Zhou as a joint PhD student of the University of Georgia.
B. K. Kendrick acknowledges that part of this work was done under the auspices of the U.S. Department of Energy under project No. 20170221ER of the Laboratory Directed Research and Development Program at Los Alamos National Laboratory. Los Alamos National Laboratory is operated by Triad National Security, LLC, for the National Nuclear Security administration of the U.S. Department of Energy (Contract No. 89233218CNA000001).
We thank Ionel Simbotin and Robin C\^ot\'e for helpful discussions.
\end{acknowledgement}

\begin{suppinfo}

Parameters used in the calculations, cross sections summed over all final states, cross sections for different partial waves, comparisons of CCI-ABC and BKMP2-APH3D results, adiabatic potentials, and the lifetime of Feshbach resonance (PDF).

\end{suppinfo}
%%%%%%%%%%%%%%%%%%%%%%%%%%%%%%%%%%%%%%%%%%%%%%%%%%%%%%%%%%%%%%%%%%%%%
%% The appropriate \bibliography command should be placed here.
%% Notice that the class file automatically sets \bibliographystyle
%% and also names the section correctly.
%%%%%%%%%%%%%%%%%%%%%%%%%%%%%%%%%%%%%%%%%%%%%%%%%%%%%%%%%%%%%%%%%%%%%
\bibliography{achemso-demo}

\providecommand{\latin}[1]{#1}
\makeatletter
\providecommand{\doi}
  {\begingroup\let\do\@makeother\dospecials
  \catcode`\{=1 \catcode`\}=2\doi@aux}
\providecommand{\doi@aux}[1]{\endgroup\texttt{#1}}
\makeatother
\providecommand*\mcitethebibliography{\thebibliography}
\csname @ifundefined\endcsname{endmcitethebibliography}
  {\let\endmcitethebibliography\endthebibliography}{}
\begin{mcitethebibliography}{43}
\providecommand*\natexlab[1]{#1}
\providecommand*\mciteSetBstSublistMode[1]{}
\providecommand*\mciteSetBstMaxWidthForm[2]{}
\providecommand*\mciteBstWouldAddEndPuncttrue
  {\def\EndOfBibitem{\unskip.}}
\providecommand*\mciteBstWouldAddEndPunctfalse
  {\let\EndOfBibitem\relax}
\providecommand*\mciteSetBstMidEndSepPunct[3]{}
\providecommand*\mciteSetBstSublistLabelBeginEnd[3]{}
\providecommand*\EndOfBibitem{}
\mciteSetBstSublistMode{f}
\mciteSetBstMaxWidthForm{subitem}{(\alph{mcitesubitemcount})}
\mciteSetBstSublistLabelBeginEnd
  {\mcitemaxwidthsubitemform\space}
  {\relax}
  {\relax}

\bibitem[Yang \latin{et~al.}(2019)Yang, Huang, Xiao, Chen, Wang, Dai, Lique,
  Alexander, Sun, and Zhang]{yang2019enhanced}
Yang,~T.; Huang,~L.; Xiao,~C.; Chen,~J.; Wang,~T.; Dai,~D.; Lique,~F.;
  Alexander,~M.~H.; Sun,~Z.; Zhang,~D. H.~{\it et al}. Enhanced reactivity of
  flourine with para-hydrogen in cold interstellar clouds by resonance-induced
  quantum tunnelling. \emph{Nat. Chem.} \textbf{2019}, \emph{11},
  744--749\relax
\mciteBstWouldAddEndPuncttrue
\mciteSetBstMidEndSepPunct{\mcitedefaultmidpunct}
{\mcitedefaultendpunct}{\mcitedefaultseppunct}\relax
\EndOfBibitem
\bibitem[Yang and Yang(2020)Yang, and Yang]{yang2020}
Yang,~T.; Yang,~X. Quantum resonances near absolute zero. \emph{Science}
  \textbf{2020}, \emph{368}, 582--583\relax
\mciteBstWouldAddEndPuncttrue
\mciteSetBstMidEndSepPunct{\mcitedefaultmidpunct}
{\mcitedefaultendpunct}{\mcitedefaultseppunct}\relax
\EndOfBibitem
\bibitem[Zhang and Miller(1989)Zhang, and Miller]{zhang1989quantum}
Zhang,~J.~Z.; Miller,~W.~H. Quantum reactive scattering via the S-matrix
  version of the Kohn variational principle: Differential and integral cross
  sections for D + H$_2$ $\to$ HD + H. \emph{J. Chem. Phys.} \textbf{1989},
  \emph{91}, 1528--1547\relax
\mciteBstWouldAddEndPuncttrue
\mciteSetBstMidEndSepPunct{\mcitedefaultmidpunct}
{\mcitedefaultendpunct}{\mcitedefaultseppunct}\relax
\EndOfBibitem
\bibitem[Takada \latin{et~al.}(1992)Takada, Ohsaki, and
  Nakamura]{takada1992reaction}
Takada,~S.; Ohsaki,~A.; Nakamura,~H. Reaction dynamics of D + H$_2$ $\to$ DH +
  H: Effects of potential energy surface topography and usefulness of the
  constant centrifugal potential approximation. \emph{J. Chem. Phys.}
  \textbf{1992}, \emph{96}, 339--348\relax
\mciteBstWouldAddEndPuncttrue
\mciteSetBstMidEndSepPunct{\mcitedefaultmidpunct}
{\mcitedefaultendpunct}{\mcitedefaultseppunct}\relax
\EndOfBibitem
\bibitem[Aoiz \latin{et~al.}(1996)Aoiz, Ba{\~n}ares, D{\'\i}ez-Rojo, Herrero,
  and S{\'a}ez~R{\'a}banos]{aoiz1996reaction}
Aoiz,~F.~J.; Ba{\~n}ares,~L.; D{\'\i}ez-Rojo,~T.; Herrero,~V.~J.;
  S{\'a}ez~R{\'a}banos,~V. Reaction cross section and rate constant
  calculations for the D + H$_2$ ($v= 0, 1$) $\to$ HD + H reaction on three
  {\it ab initio} potential energy surfaces. A quasiclassical trajectory study.
  \emph{J. Phys. Chem.} \textbf{1996}, \emph{100}, 4071--4083\relax
\mciteBstWouldAddEndPuncttrue
\mciteSetBstMidEndSepPunct{\mcitedefaultmidpunct}
{\mcitedefaultendpunct}{\mcitedefaultseppunct}\relax
\EndOfBibitem
\bibitem[Banares and D'Mello(1997)Banares, and D'Mello]{banares1997quantum}
Banares,~L.; D'Mello,~M. Quantum mechanical rate constants for the D + H$_2$
  $\to$ HD + H reaction on the BKMP2 potential energy surface. \emph{Chem.
  Phys. Lett.} \textbf{1997}, \emph{277}, 465--472\relax
\mciteBstWouldAddEndPuncttrue
\mciteSetBstMidEndSepPunct{\mcitedefaultmidpunct}
{\mcitedefaultendpunct}{\mcitedefaultseppunct}\relax
\EndOfBibitem
\bibitem[Yuan \latin{et~al.}(2020)Yuan, Chen, Luo, Tan, Li, Huang, Sun, Yang,
  and Wang]{yuan2020imaging}
Yuan,~D.; Chen,~W.; Luo,~C.; Tan,~Y.; Li,~S.; Huang,~Y.; Sun,~Z.; Yang,~X.;
  Wang,~X. Imaging the State-to-State Dynamics of the H+ D$_2$ $\to$ HD + D
  Reaction at 1.42 eV. \emph{J. Phys. Chem. Lett.} \textbf{2020}, \emph{11},
  1222--1227\relax
\mciteBstWouldAddEndPuncttrue
\mciteSetBstMidEndSepPunct{\mcitedefaultmidpunct}
{\mcitedefaultendpunct}{\mcitedefaultseppunct}\relax
\EndOfBibitem
\bibitem[Marinero \latin{et~al.}(1984)Marinero, Rettner, and
  Zare]{marinero1984h+}
Marinero,~E.~E.; Rettner,~C.~T.; Zare,~R.~N. H + D$_2$ reaction dynamics.
  Determination of the product state distributions at a collision energy of 1.3
  eV. \emph{J. Chem. Phys.} \textbf{1984}, \emph{80}, 4142--4156\relax
\mciteBstWouldAddEndPuncttrue
\mciteSetBstMidEndSepPunct{\mcitedefaultmidpunct}
{\mcitedefaultendpunct}{\mcitedefaultseppunct}\relax
\EndOfBibitem
\bibitem[Rinnen \latin{et~al.}(1988)Rinnen, Kliner, Blake, and
  Zare]{rinnen1988h+}
Rinnen,~K.-D.; Kliner,~D.~A.; Blake,~R.~S.; Zare,~R.~N. The H + D$_2$ reaction:
  ``prompt'' HD distributions at high collision energies. \emph{Chem. Phys.
  Lett.} \textbf{1988}, \emph{153}, 371--375\relax
\mciteBstWouldAddEndPuncttrue
\mciteSetBstMidEndSepPunct{\mcitedefaultmidpunct}
{\mcitedefaultendpunct}{\mcitedefaultseppunct}\relax
\EndOfBibitem
\bibitem[D'Mello \latin{et~al.}(1991)D'Mello, Manolopoulos, and
  Wyatt]{d1991quantum}
D'Mello,~M.; Manolopoulos,~D.~E.; Wyatt,~R.~E. Quantum dynamics of the H +
  D$_2$ $\to$ D + HD reaction: Comparison with experiment. \emph{J. Chem.
  Phys.} \textbf{1991}, \emph{94}, 5985--5993\relax
\mciteBstWouldAddEndPuncttrue
\mciteSetBstMidEndSepPunct{\mcitedefaultmidpunct}
{\mcitedefaultendpunct}{\mcitedefaultseppunct}\relax
\EndOfBibitem
\bibitem[Gao \latin{et~al.}(2015)Gao, Sneha, Bouakline, Althorpe, and
  Zare]{gao2015differential}
Gao,~H.; Sneha,~M.; Bouakline,~F.; Althorpe,~S.~C.; Zare,~R.~N. Differential
  Cross Sections for the H + D$_2$ $\to$ HD ($v'= 3, j'= 4-10$) + D reaction
  above the Conical Intersection. \emph{J. Phys. Chem. A} \textbf{2015},
  \emph{119}, 12036--12042\relax
\mciteBstWouldAddEndPuncttrue
\mciteSetBstMidEndSepPunct{\mcitedefaultmidpunct}
{\mcitedefaultendpunct}{\mcitedefaultseppunct}\relax
\EndOfBibitem
\bibitem[Yuan \latin{et~al.}(2018)Yuan, Yu, Chen, Sang, Luo, Wang, Xu,
  Casavecchia, Wang, and Sun]{yuan2018direct}
Yuan,~D.; Yu,~S.; Chen,~W.; Sang,~J.; Luo,~C.; Wang,~T.; Xu,~X.;
  Casavecchia,~P.; Wang,~X.; Sun,~Z.~{\it et al}. Direct observation of
  forward-scattering oscillations in the H + HD $\to$ H$_2$ + D reaction.
  \emph{Nat. Chem.} \textbf{2018}, \emph{10}, 653--658\relax
\mciteBstWouldAddEndPuncttrue
\mciteSetBstMidEndSepPunct{\mcitedefaultmidpunct}
{\mcitedefaultendpunct}{\mcitedefaultseppunct}\relax
\EndOfBibitem
\bibitem[Yuan \latin{et~al.}(2018)Yuan, Guan, Chen, Zhao, Yu, Luo, Tan, Xie,
  Wang, and Sun]{yuan2018observation}
Yuan,~D.; Guan,~Y.; Chen,~W.; Zhao,~H.; Yu,~S.; Luo,~C.; Tan,~Y.; Xie,~T.;
  Wang,~X.; Sun,~Z.~{\it et al}. Observation of the geometric phase effect in
  the H + HD $\to$ H$_2$ + D reaction. \emph{Science} \textbf{2018},
  \emph{362}, 1289--1293\relax
\mciteBstWouldAddEndPuncttrue
\mciteSetBstMidEndSepPunct{\mcitedefaultmidpunct}
{\mcitedefaultendpunct}{\mcitedefaultseppunct}\relax
\EndOfBibitem
\bibitem[Desrousseaux \latin{et~al.}(2018)Desrousseaux, Coppola, Kazandjian,
  and Lique]{desrousseaux2018rotational}
Desrousseaux,~B.; Coppola,~C.~M.; Kazandjian,~M.~V.; Lique,~F. Rotational
  Excitation of HD by Hydrogen Revisited. \emph{J. Phys. Chem. A}
  \textbf{2018}, \emph{122}, 8390--8396\relax
\mciteBstWouldAddEndPuncttrue
\mciteSetBstMidEndSepPunct{\mcitedefaultmidpunct}
{\mcitedefaultendpunct}{\mcitedefaultseppunct}\relax
\EndOfBibitem
\bibitem[Kendrick \latin{et~al.}(2016)Kendrick, Hazra, and
  Balakrishnan]{kendrick2016geometric1}
Kendrick,~B.~K.; Hazra,~J.; Balakrishnan,~N. Geometric phase effects in the
  ultracold D + HD $\to$ D + HD and D + HD $\to$ H + D$_2$ reactions. \emph{New
  J. Phys} \textbf{2016}, \emph{18}, 123020\relax
\mciteBstWouldAddEndPuncttrue
\mciteSetBstMidEndSepPunct{\mcitedefaultmidpunct}
{\mcitedefaultendpunct}{\mcitedefaultseppunct}\relax
\EndOfBibitem
\bibitem[Kendrick \latin{et~al.}(2016)Kendrick, Hazra, and
  Balakrishnan]{kendrick2016geometric2}
Kendrick,~B.~K.; Hazra,~J.; Balakrishnan,~N. Geometric phase effects in the
  ultracold H + H$_2$ reaction. \emph{J. Chem. Phys.} \textbf{2016},
  \emph{145}, 164303\relax
\mciteBstWouldAddEndPuncttrue
\mciteSetBstMidEndSepPunct{\mcitedefaultmidpunct}
{\mcitedefaultendpunct}{\mcitedefaultseppunct}\relax
\EndOfBibitem
\bibitem[Hazra \latin{et~al.}(2016)Hazra, Kendrick, and
  Balakrishnan]{hazra2016geometric}
Hazra,~J.; Kendrick,~B.~K.; Balakrishnan,~N. Geometric phase effects in
  ultracold hydrogen exchange reaction. \emph{J. Phys. B: At., Mol. Opt. Phys.}
  \textbf{2016}, \emph{49}, 194004\relax
\mciteBstWouldAddEndPuncttrue
\mciteSetBstMidEndSepPunct{\mcitedefaultmidpunct}
{\mcitedefaultendpunct}{\mcitedefaultseppunct}\relax
\EndOfBibitem
\bibitem[Kendrick \latin{et~al.}(2015)Kendrick, Hazra, and
  Balakrishnan]{kendrick2015geometric}
Kendrick,~B.~K.; Hazra,~J.; Balakrishnan,~N. Geometric phase appears in the
  ultracold hydrogen exchange reaction. \emph{Phys. Rev. Lett.} \textbf{2015},
  \emph{115}, 153201\relax
\mciteBstWouldAddEndPuncttrue
\mciteSetBstMidEndSepPunct{\mcitedefaultmidpunct}
{\mcitedefaultendpunct}{\mcitedefaultseppunct}\relax
\EndOfBibitem
\bibitem[Kendrick(2019)]{kendrick2019nonadiabatic}
Kendrick,~B.~K. Nonadiabatic Ultracold Quantum Reactive Scattering of Hydrogen
  with Vibrationally Excited HD ($v= 5-9$). \emph{J. Phys. Chem. A}
  \textbf{2019}, \emph{123}, 9919--9933\relax
\mciteBstWouldAddEndPuncttrue
\mciteSetBstMidEndSepPunct{\mcitedefaultmidpunct}
{\mcitedefaultendpunct}{\mcitedefaultseppunct}\relax
\EndOfBibitem
\bibitem[Perreault \latin{et~al.}(2017)Perreault, Mukherjee, and
  Zare]{perreault2017quantum}
Perreault,~W.~E.; Mukherjee,~N.; Zare,~R.~N. Quantum control of molecular
  collisions at 1 kelvin. \emph{Science} \textbf{2017}, \emph{358},
  356--359\relax
\mciteBstWouldAddEndPuncttrue
\mciteSetBstMidEndSepPunct{\mcitedefaultmidpunct}
{\mcitedefaultendpunct}{\mcitedefaultseppunct}\relax
\EndOfBibitem
\bibitem[Perreault \latin{et~al.}(2018)Perreault, Mukherjee, and
  Zare]{perreault2018cold}
Perreault,~W.~E.; Mukherjee,~N.; Zare,~R.~N. Cold quantum-controlled
  rotationally inelastic scattering of HD with H$_2$ and D$_2$ reveals
  collisional partner reorientation. \emph{Nat. Chem.} \textbf{2018},
  \emph{10}, 561--567\relax
\mciteBstWouldAddEndPuncttrue
\mciteSetBstMidEndSepPunct{\mcitedefaultmidpunct}
{\mcitedefaultendpunct}{\mcitedefaultseppunct}\relax
\EndOfBibitem
\bibitem[Croft \latin{et~al.}(2018)Croft, Balakrishnan, Huang, and
  Guo]{croft2018unraveling}
Croft,~J.~F.; Balakrishnan,~N.; Huang,~M.; Guo,~H. Unraveling the
  Stereodynamics of Cold Controlled HD--H$_2$ Collisions. \emph{Phys. Rev.
  Lett.} \textbf{2018}, \emph{121}, 113401\relax
\mciteBstWouldAddEndPuncttrue
\mciteSetBstMidEndSepPunct{\mcitedefaultmidpunct}
{\mcitedefaultendpunct}{\mcitedefaultseppunct}\relax
\EndOfBibitem
\bibitem[Croft and Balakrishnan(2019)Croft, and
  Balakrishnan]{croft2019controlling}
Croft,~J.~F.; Balakrishnan,~N. Controlling rotational quenching rates in cold
  molecular collisions. \emph{J. Chem. Phys.} \textbf{2019}, \emph{150},
  164302\relax
\mciteBstWouldAddEndPuncttrue
\mciteSetBstMidEndSepPunct{\mcitedefaultmidpunct}
{\mcitedefaultendpunct}{\mcitedefaultseppunct}\relax
\EndOfBibitem
\bibitem[Perreault \latin{et~al.}(2019)Perreault, Mukherjee, and
  Zare]{perreault2019hd}
Perreault,~W.~E.; Mukherjee,~N.; Zare,~R.~N. HD ($v= 1, j= 2, m$) orientation
  controls HD--He rotationally inelastic scattering near 1 K. \emph{J. Chem.
  Phys.} \textbf{2019}, \emph{150}, 174301\relax
\mciteBstWouldAddEndPuncttrue
\mciteSetBstMidEndSepPunct{\mcitedefaultmidpunct}
{\mcitedefaultendpunct}{\mcitedefaultseppunct}\relax
\EndOfBibitem
\bibitem[Liu(1973)]{liu1973ab}
Liu,~B. Ab initio potential energy surface for linear H$_3$. \emph{J. Chem.
  Phys.} \textbf{1973}, \emph{58}, 1925--1937\relax
\mciteBstWouldAddEndPuncttrue
\mciteSetBstMidEndSepPunct{\mcitedefaultmidpunct}
{\mcitedefaultendpunct}{\mcitedefaultseppunct}\relax
\EndOfBibitem
\bibitem[Siegbahn and Liu(1978)Siegbahn, and Liu]{siegbahn1978accurate}
Siegbahn,~P.; Liu,~B. An accurate three-dimensional potential energy surface
  for H$_3$. \emph{J. Chem. Phys.} \textbf{1978}, \emph{68}, 2457--2465\relax
\mciteBstWouldAddEndPuncttrue
\mciteSetBstMidEndSepPunct{\mcitedefaultmidpunct}
{\mcitedefaultendpunct}{\mcitedefaultseppunct}\relax
\EndOfBibitem
\bibitem[Truhlar and Horowitz(1978)Truhlar, and
  Horowitz]{truhlar1978functional}
Truhlar,~D.~G.; Horowitz,~C.~J. Functional representation of Liu and
  Siegbahn’s accurate abinitio potential energy calculations for H + H$_2$.
  \emph{J. Chem. Phys.} \textbf{1978}, \emph{68}, 2466--2476\relax
\mciteBstWouldAddEndPuncttrue
\mciteSetBstMidEndSepPunct{\mcitedefaultmidpunct}
{\mcitedefaultendpunct}{\mcitedefaultseppunct}\relax
\EndOfBibitem
\bibitem[Boothroyd \latin{et~al.}(1996)Boothroyd, Keogh, Martin, and
  Peterson]{boothroyd1996refined}
Boothroyd,~A.~I.; Keogh,~W.~J.; Martin,~P.~G.; Peterson,~M.~R. A refined H$_3$
  potential energy surface. \emph{J. Chem. Phys.} \textbf{1996}, \emph{104},
  7139--7152\relax
\mciteBstWouldAddEndPuncttrue
\mciteSetBstMidEndSepPunct{\mcitedefaultmidpunct}
{\mcitedefaultendpunct}{\mcitedefaultseppunct}\relax
\EndOfBibitem
\bibitem[Mielke \latin{et~al.}(2002)Mielke, Garrett, and
  Peterson]{mielke2002hierarchical}
Mielke,~S.~L.; Garrett,~B.~C.; Peterson,~K.~A. A hierarchical family of global
  analytic Born--Oppenheimer potential energy surfaces for the H + H$_2$
  reaction ranging in quality from double-zeta to the complete basis set limit.
  \emph{J. Chem. Phys.} \textbf{2002}, \emph{116}, 4142--4161\relax
\mciteBstWouldAddEndPuncttrue
\mciteSetBstMidEndSepPunct{\mcitedefaultmidpunct}
{\mcitedefaultendpunct}{\mcitedefaultseppunct}\relax
\EndOfBibitem
\bibitem[Skouteris \latin{et~al.}(2000)Skouteris, Castillo, and
  Manolopoulos]{skouteris2000abc}
Skouteris,~D.; Castillo,~J.; Manolopoulos,~D. ABC: a quantum reactive
  scattering program. \emph{Comput. Phys. Commun.} \textbf{2000}, \emph{133},
  128--135\relax
\mciteBstWouldAddEndPuncttrue
\mciteSetBstMidEndSepPunct{\mcitedefaultmidpunct}
{\mcitedefaultendpunct}{\mcitedefaultseppunct}\relax
\EndOfBibitem
\bibitem[Der~Chao and Skodje(2002)Der~Chao, and Skodje]{der2002signatures}
Der~Chao,~S.; Skodje,~R.~T. Signatures of reactive resonance: three case
  studies. \emph{Theor. Chem. Acc.} \textbf{2002}, \emph{108}, 273--285\relax
\mciteBstWouldAddEndPuncttrue
\mciteSetBstMidEndSepPunct{\mcitedefaultmidpunct}
{\mcitedefaultendpunct}{\mcitedefaultseppunct}\relax
\EndOfBibitem
\bibitem[Kendrick and Pack(1995)Kendrick, and Pack]{kendrick1995recombination}
Kendrick,~B.; Pack,~R.~T. Recombination resonances in thermal H + O$_2$
  scattering. \emph{Chem. Phys. Lett.} \textbf{1995}, \emph{235},
  291--296\relax
\mciteBstWouldAddEndPuncttrue
\mciteSetBstMidEndSepPunct{\mcitedefaultmidpunct}
{\mcitedefaultendpunct}{\mcitedefaultseppunct}\relax
\EndOfBibitem
\bibitem[Kendrick(2003)]{kendrick2003quantum}
Kendrick,~B.~K. Quantum reactive scattering calculations for the D + H$_2$
  $\to$ HD + H reaction. \emph{J. Chem. Phys.} \textbf{2003}, \emph{118},
  10502--10522\relax
\mciteBstWouldAddEndPuncttrue
\mciteSetBstMidEndSepPunct{\mcitedefaultmidpunct}
{\mcitedefaultendpunct}{\mcitedefaultseppunct}\relax
\EndOfBibitem
\bibitem[Smith(1960)]{smith1960lifetime}
Smith,~F.~T. Lifetime matrix in collision theory. \emph{Phys. Rev.}
  \textbf{1960}, \emph{118}, 349--356\relax
\mciteBstWouldAddEndPuncttrue
\mciteSetBstMidEndSepPunct{\mcitedefaultmidpunct}
{\mcitedefaultendpunct}{\mcitedefaultseppunct}\relax
\EndOfBibitem
\bibitem[Mukherjee \latin{et~al.}(2018)Mukherjee, Perreault, and
  Zare]{mukherjee2018stark}
Mukherjee,~N.; Perreault,~W.~E.; Zare,~R.~N. \emph{Frontiers and Advances in
  Molecular Spectroscopy}; Elsevier: Amsterdam, NL, 2018; pp 1--46\relax
\mciteBstWouldAddEndPuncttrue
\mciteSetBstMidEndSepPunct{\mcitedefaultmidpunct}
{\mcitedefaultendpunct}{\mcitedefaultseppunct}\relax
\EndOfBibitem
\bibitem[Harich \latin{et~al.}(2002)Harich, Dai, Wang, Yang, Der~Chao, and
  Skodje]{harich2002forward}
Harich,~S.~A.; Dai,~D.; Wang,~C.~C.; Yang,~X.; Der~Chao,~S.; Skodje,~R.~T.
  Forward scattering due to slow-down of the intermediate in the H + HD $\to$ D
  + H$_2$ reaction. \emph{Nature} \textbf{2002}, \emph{419}, 281--284\relax
\mciteBstWouldAddEndPuncttrue
\mciteSetBstMidEndSepPunct{\mcitedefaultmidpunct}
{\mcitedefaultendpunct}{\mcitedefaultseppunct}\relax
\EndOfBibitem
\bibitem[Dai \latin{et~al.}(2003)Dai, Wang, Harich, Wang, Yang, Der~Chao, and
  Skodje]{dai2003interference}
Dai,~D.; Wang,~C.~C.; Harich,~S.~A.; Wang,~X.; Yang,~X.; Der~Chao,~S.;
  Skodje,~R.~T. Interference of quantized transition-state pathways in the H +
  D$_2$ $\to$ D + HD chemical reaction. \emph{Science} \textbf{2003},
  \emph{300}, 1730--1734\relax
\mciteBstWouldAddEndPuncttrue
\mciteSetBstMidEndSepPunct{\mcitedefaultmidpunct}
{\mcitedefaultendpunct}{\mcitedefaultseppunct}\relax
\EndOfBibitem
\bibitem[Jankunas \latin{et~al.}(2012)Jankunas, Zare, Bouakline, Althorpe,
  Herr{\'a}ez-Aguilar, and Aoiz]{jankunas2012seemingly}
Jankunas,~J.; Zare,~R.~N.; Bouakline,~F.; Althorpe,~S.~C.;
  Herr{\'a}ez-Aguilar,~D.; Aoiz,~F.~J. Seemingly anomalous angular
  distributions in H + D$_2$ reactive scattering. \emph{Science} \textbf{2012},
  \emph{336}, 1687--1690\relax
\mciteBstWouldAddEndPuncttrue
\mciteSetBstMidEndSepPunct{\mcitedefaultmidpunct}
{\mcitedefaultendpunct}{\mcitedefaultseppunct}\relax
\EndOfBibitem
\bibitem[Jankunas \latin{et~al.}(2014)Jankunas, Sneha, Zare, Bouakline,
  Althorpe, Herr{\'a}ez-Aguilar, and Aoiz]{jankunas2014simplest}
Jankunas,~J.; Sneha,~M.; Zare,~R.~N.; Bouakline,~F.; Althorpe,~S.~C.;
  Herr{\'a}ez-Aguilar,~D.; Aoiz,~F.~J. Is the simplest chemical reaction really
  so simple? \emph{Proc. Natl. Acad. Sci. U. S. A.} \textbf{2014}, \emph{111},
  15--20\relax
\mciteBstWouldAddEndPuncttrue
\mciteSetBstMidEndSepPunct{\mcitedefaultmidpunct}
{\mcitedefaultendpunct}{\mcitedefaultseppunct}\relax
\EndOfBibitem
\bibitem[Eppink and Parker(1997)Eppink, and Parker]{eppink1997velocity}
Eppink,~A.~T.; Parker,~D.~H. Velocity map imaging of ions and electrons using
  electrostatic lenses: Application in photoelectron and photofragment ion
  imaging of molecular oxygen. \emph{Rev. Sci. Instrum.} \textbf{1997},
  \emph{68}, 3477--3484\relax
\mciteBstWouldAddEndPuncttrue
\mciteSetBstMidEndSepPunct{\mcitedefaultmidpunct}
{\mcitedefaultendpunct}{\mcitedefaultseppunct}\relax
\EndOfBibitem
\bibitem[Lin \latin{et~al.}(2003)Lin, Zhou, Shiu, and Liu]{lin2003application}
Lin,~J.~J.; Zhou,~J.; Shiu,~W.; Liu,~K. Application of time-sliced ion velocity
  imaging to crossed molecular beam experiments. \emph{Rev. Sci. Instrum.}
  \textbf{2003}, \emph{74}, 2495--2500\relax
\mciteBstWouldAddEndPuncttrue
\mciteSetBstMidEndSepPunct{\mcitedefaultmidpunct}
{\mcitedefaultendpunct}{\mcitedefaultseppunct}\relax
\EndOfBibitem
\bibitem[Zimmermann \latin{et~al.}(2008)Zimmermann, M{\"u}hlberger, Fuhrer,
  Gonin, and Welthagen]{zimmermann2008ultracompact}
Zimmermann,~R.; M{\"u}hlberger,~F.; Fuhrer,~K.; Gonin,~M.; Welthagen,~W. An
  ultracompact photo-ionization time-of-flight mass spectrometer with a novel
  vacuum ultraviolet light source for on-line detection of organic trace
  compounds and as a detector for gas chromatography. \emph{J. Mater. Cycles
  Waste Manage.} \textbf{2008}, \emph{10}, 24--31\relax
\mciteBstWouldAddEndPuncttrue
\mciteSetBstMidEndSepPunct{\mcitedefaultmidpunct}
{\mcitedefaultendpunct}{\mcitedefaultseppunct}\relax
\EndOfBibitem
\end{mcitethebibliography}

\end{document}